\documentclass[11pt,letterpaper]{JHEP3}

\usepackage{epsfig,multicol,bbm}
\usepackage{amsfonts}
\usepackage{amsmath,amssymb,epsfig}
\usepackage{mqcommands}

\newcommand{\nn}{\mathcal N(r)}
\newcommand{\oef}{\Omega_{\mbox{\tiny{eff}}}}

\newcommand{\reef}[1]{(\ref{#1})}
\newcommand{\ff}{f_{\infty}}
\renewcommand{\eqref}[1]{(\ref{#1})}

\newcommand{\lp}{l_{p}}

\title{Holographic phase space: $c$-functions and black holes as renormalization group flows.}
\preprint{}
\author{Miguel F. Paulos$^{a}$\\
$^a$ {\it Laboratoire de Physique Th\'eorique et Hautes Energies, CNRS UMR 7589,\\ Universit\'e Pierre et Marie Curie, 4 place Jussieu, 75252 Paris Cedex 05, France }\\

\vskip .5cm

{\rm E-mail:}\ \ {\
mpaulos@lpthe.jussieu.fr}}

\abstract{We construct a $\mathcal N$-function for Lovelock theories of gravity, which yields a holographic $c$-function in domain-wall backgrounds, and seemingly generalizes the concept for black hole geometries. A flow equation equates the monotonicity properties of  $\mathcal N$ with the gravitational field, which has opposite signs in the domain-wall and black hole backgrounds, due to the presence of negative/positive energy in the former/latter, and accordingly $\mathcal N$ monotonically decreases/increases from the UV to the IR.
On $AdS$ spaces the $\mathcal N$-function is related to the Euler anomaly, and at a black hole horizon it is generically proportional to the entropy. For planar black holes, $\mathcal N$ diverges at the horizon, which we interpret as an order $N^2$ increase in the number of effective degrees of freedom.  We show how $\mathcal N$ can be written as the ratio of the Wald entropy to an effective phase space volume, and using the flow equation relate this to Verlinde's notion of gravity as an entropic force. From the effective phase space we can obtain an expression for the dual field theory momentum cut-off, matching a previous proposal in the literature by Polchinski and Heemskerk. Finally, we propose that the area in Planck units counts states, not degrees of freedom, and identify it also as a phase space volume.  Written in terms of the proper radial distance $\beta$, it takes the suggestive form of a canonical partition function at inverse temperature $\beta$, leading to a ``mean energy'' which is simply the extrinsic curvature of the surface. Using this we relate this definition of holographic phase space with the effective phase space appearing in the $\mathcal N$-function.

}


\begin{document}{\vskip 1cm}

\section{Introduction}

In two dimensional conformal field theories (CFTs), the number of degrees of freedom can be equated with the central charge of the theory $c$, and its running with scale is described by a $c$-function \cite{Zamolodchikov:1986gt}. This is a function which is monotonously decreasing under renormalization group (RG) flow and which counts number of degrees of freedom in the sense that when evaluated at ultra-violet (UV) and infra-red (IR) conformal fixed points it yields the central charge of the theory. Monotonocity then implies $c_{\mbox{\tiny{UV}}}\geq c_{\mbox{\tiny{IR}}}$.  While this story is well understood in two dimensions, a proof of existence of a $c$-function in higher dimensional CFTs is not available. Part of the problem is that in higher dimensional conformal field theories there are several candidates for replacing the role played by $c$ in two dimensions. It was Cardy \cite{Cardy:1988cwa} who originally suggested that at least for even-dimensional conformal field theories, the appropriate quantity to consider is $A$, or Euler anomaly - essentially the prefactor multiplying the Euler density $E_d$ in the Weyl anomaly of the theory, which looks schematically as
\be
\langle T^{a}_{\ a}\rangle \simeq (-1)^{d/2} A E_{d}+\sum B_i C_i^{d/2}+\mbox{total derivatives},
\ee
with $C$ the Weyl tensor, and $C_i^{d/2}$ a set of independent scalars built out of $d/2$ Weyl tensors.
Recently, evidence for this proposal has been found in the context of the $AdS$/CFT correspondence  \cite{Maldacena:1997re,Aharony:1999ti}. From a holographic perspective it is relatively straightforward to construct a $c$-function for gravitational theories described by Einstein gravity in $AdS$ space \cite{Girardello:1998pd,Freedman:1999gp}. Unfortunately, Einstein gravity is too simple, and it is not possible to verify what exactly this $c$-function matches onto holographically, since all holographic anomalies are essentially the same in this case. In order to distinguish these one must consider more complicated gravity theories (for some early work on this see \cite{Nojiri:1999mh}). By examinining holographic c-functions in a particular set of gravitational theories\footnote{Note also that conversely, demanding that such a function exists can serve as a useful constraint for building interesting theories \cite{Sinha:2010ai,Sinha:2010pm,Paulos:2010ke}. }, the work in \cite{Myers:2010xs} suggests that the natural quantity to consider is indeed $A$, or a certain coefficient appearing in entanglement entropy between two-halves of a sphere, for odd-dimensional conformal field theories. 

One way of thinking about holographic $c$-functions, is that they provide a notion of what are the number of degrees of freedom that are active at a given scale, where scale is identified with the radial coordinate in $AdS$ space. But, a seemingly different point of view on such degrees of freedom was provided in \cite{Susskind:1998dq}, which argued that the number of degrees of freedom which holographically describe a given spacetime region should be proportional to the area in Planck units. Such a proposal leads to the right number of degrees of freedom for $N=4$ supersymmetric Yang-Mills theory at strong coupling. More recently, \cite{Padmanabhan:2010xh} has shown that gravitational equations of motion describing  a closed region of spacetime can be recast as thermodynamic relations on its surface. For higher derivative theories of gravity, the ``surface density of degrees of freedom'' holographically describing this region, is directly related to the Wald formula \cite{Iyer:1994ys,Wald:1993nt} for black hole entropy. It is not immediately obvious how these different proposals for the number of degrees of freedom relate to the holographic c-functions.

Another interesting question which has for the moment not been explored, is the relation between $c$-functions and black hole entropy. There are many hints that such a connection might exist. As mentioned above, the association of one degree of freedom per Planck area leads to the correct state counting, as captured by central charges or Weyl anomalies of the dual CFT, but it also leads to the correct value for black hole entropy, a setting which is seemingly far removed from the concept of $c$-functions, which describe RG flows in the vacuum state. There is indeed evidence that asymptotically $AdS$ black hole solutions describe a new, intriguing type of renormalization group flow, between conformal field theories of different dimensionalities. Perhaps the clearest case for this is obtained by considering extremal Reissner-N\"ordstrom black holes in $AdS$, which have a near horizon $AdS_2$ factor associated with an emergent IR conformal field theory \cite{Faulkner:2009wj}; but even in non-extremal black holes there seems to be an emergent conformal symmetry at the horizon \cite{Solodukhin:1998tc,Carlip:2008rk}. These geometries capture renormalization group flow in a field theory which is in a non-trivial thermal state, or a state at finite density. As such, our usual intuition on $c$-functions might not hold here in these cases.

In this paper, we will attempt an initial exploration of these questions, by considering holographic $c$-functions and their generalization in black hole backgrounds, in the context of Lovelock theories of gravity \cite{Lovelock1971,Zumino1986}. Our choice to focus on these theories stems from the fact that they have enough structure to distinguish between various kinds of central charges, or other quantities which would be degenerate in Einstein gravity; and yet are still amenable to a simple analytic treatment. The downside is that no known string theory embedding of these theories, although these theories seem to holographically capture a lot of interesting features of a realistic conformal field theory \cite{Hofman:2009ug,Buchel:2009sk,Camanho:2010ru}.
By examining the gravitational equations of motion in a class of black hole backgrounds, we will find that these can be recast as a flow equation for what we will denote as the $\mathcal N$-function. This equation tells us how the $\mathcal N$-function varies depending on the local ``gravitational field''. For domain-wall backgrounds describing vacuum RG flows, $\mathcal N$ directly provides a holographic $c$-function, for any Lovelock theory. 

For black hole backgrounds the story is more interesting. Depending on the horizon topology, the behaviour of $\mathcal N$ changes, although it is always  monotonously increasing towards the horizon for vacuum black hole solutions, thus directly contradicting our usual expectations for the monotonicity of a $c$-function. We shall see that this directly follows from the flow equation, and from the fact that the gravitational field points in opposite directions in domain-wall solutions, where there is negative energy density, and in black hole solutions where there is positive energy density. We shall find that in any case, $\mathcal N$ evaluated at the horizon can be directly related to the black hole entropy. A puzzle appears however, since $\mathcal N$ can be divergent, either at the horizon for the planar black hole, or  before it for the spherical one. Softening the divergence by introducing a Planck scale cut-off, then in the planar black hole case we see that $\mathcal N$ increases dramatically from the boundary to the horizon by a factor of $(L/\lp)^{d-2}\simeq N^2$ for gauge theories at large $N$. Alternatively, this divergence could be interpreted as a signal of dimensional reduction, where spatial modes of a single field are effectively transforming into a new set of decoupled fields. However, it is hard to make sense of this divergence in the spherical case. We were unable to solve this puzzle in this work, but we offer some speculations in the discussion section.

The $\mathcal N$-function can be written in a simple fashion as the ratio of an entropy to a certain quantity that we shall interpret as an effective phase space volume $\oef$,
\be
\mathcal N=\frac{S}{4\pi \oef}.
\ee
More concretely, the first is defined as Wald's formula evaluated on a radial slice, or holographic screen, 
\be
S=-2\pi \oint \sqrt{h}\, \frac{\delta \mathcal L}{\delta R_{abcd}}\, \epsilon_{ab}\epsilon_{cd} 
\ee
and is argued to compute the total information content of the region beyond the screen: the entropy associated with the ignorance about which state is hidden behind the scren. As for the effective phase space it can be written in the form
\bea
\oef &=& \left(\frac{L}{\lp}\right)^{d-2}\oint d^{d-2}x \sqrt{h}\,\sqrt{\mbox{det}\left( K_i^k K_k^j-\,\kappa \,\delta_i^j \frac{L^2}{r^2}\right)}, \eea
with $\kappa$ the curvature of the transverse $d-2$ space. From the field theory point of view this looks like $N^2 V_{d-2} (E^2-m^2)^{(d-2)/2}$, which indeed counts the number of states accessible to a single particle excitation. The cut-off, or momentum scale is associated with $K_{ab}$ the extrinsic curvature, which is also the momentum of the metric associated with radial translations. In domain-wall backgrounds the resulting expression precisely matches a proposal for the field theory cut-off given in \cite{Heemskerk:2010hk}.

In this paper we will also attempt to relate this definition of phase space with the notion of one degree of freedom per Planck area. More concretely, we define an alternative version of phase space $\Omega$, interpreting the area in Planck units as counting states, not degrees of freedom:
\be
\Omega=\oint_{\partial M}\!\frac{dA}{\lp^{d-2}},
\ee
and accordingly an alternative notion of degrees of freedom based on this
\be
N_{\mbox{\tiny{dof}}}=\frac{S}{4\pi\Omega}.
\ee
This exactly matches the proposal for the surface density of degrees of freedom given in \cite{Padmanabhan:2010xh}. However, the above does not satisfy any simple equation of motion, unlike $\mathcal N$. In the case where $d-2$ dimensional transverse geometry is flat, there is a simple relation between the two functions,
\be
\mathcal N=\frac{N_{\mbox{\tiny{dof}}}}{(L\,\mathcal E)^{d-2}}, \qquad \mathcal E=\frac{d\log{\Omega}}{d\beta}.
\ee
Here $\beta$ is the proper radial distance. We have chosen to call it $\beta$ because, when written in terms of this quantity, the phase space volume $\Omega$ looks exactly like a canonical partition function at inverse temperature $\beta$. In fact, taking this analogy further, we can compute the mean energy and energy squared finding
\be
\langle E\rangle=- \frac{d \log \Omega}{d \beta}\simeq K, \qquad \langle E^2 \rangle\equiv \frac{1}{\Omega} \frac{\partial^2 \Omega}{\partial \beta^2}\simeq R_{abcd}h^{ac}n^{b}n^{d}.
\ee
with $h_{ab}$ the metric of a co-dimension two space-like surface, $n^a$ a vector pointing in the normal, spacelike direction to the surface, and $K=h^{ab}K_{ab}=\nabla_{a} n^a$. This is suggestive of a possible thermodynamic interpretation of the $AdS$ geometry, and we offer some speculations on this in in section \ref{thermoent}. 


This paper is structured as follows: in section \ref{sec2}, we briefly describe the set of theories we will be considering and some of their solutions. In section \ref{sec3}, by analogy with the Einstein gravity case, we define the $\mathcal N$-function, In particular we show how the evolution of $\mathcal N$ is determined by the local gravitational field, and how this is quite different in RG flow backgrounds and black hole backgrounds. In the former, $\mathcal N$ directly provides a $c$-function for any Lovelock theory. We then analyse the behaviour of $\mathcal N$ for black holes with different horizon topology and find that, when evaluated at the horizon, $\mathcal N$ is directly related to the entropy. In section \ref{sec4} we show how $\mathcal N$ can be written as the ratio of Wald information to a quantity which we interpret as an effective phase space volume, and comment on the relation of this result with some of Verlinde's proposals \cite{Verlinde:2010hp}. Section \ref{sec5} proposes that the area in Planck units provides alternative definition of phase space volume. Using the proper radial distance $\beta$ leads to intriguing similarities of the $AdS$-geometries with a thermodynamical system. We speculate on this subject in section \ref{thermoent}, where we also connect this proposal of phase space with the effective phase space appearing in the $\mathcal N$ function. We finish with a discussion and directions for future work.

{\bf Note added:} While this work was being finished, the papers \cite{Myers:2010tj,Liu:2010xc} appeared, whose results partially overlap with ours. In particular, the use of the Wald formula for the computation of the Euler anomaly $A$ which we do in appendix \ref{Euler} has appeared in \cite{Myers:2010tj}, who also defined a holographic $c$-function for Gauss-Bonnet theory in any dimension. Also \cite{Liu:2010xc} has shown how to construct c-functions for any Lovelock theory.

\section{Lovelock theories of gravity}
\label{sec2}
Throughout most of this paper, we will be interested in the gravitational dynamics of Lovelock theories in a certain class of backgrounds. As explained in the introduction, Lovelock theories are simple enough to be studied in detail, yet contain a lot more structure than Einstein gravity. A brief review of these theories and some of their elementary properties is given in appendix \reef{Lovelock}. We will admit the existence of a negative cosmological constant, and as such in general such theories can support stable $AdS$ vacua. The action is of the schematic form
\be
I_g=\frac{1}{\lp^{d-2}}\int d^{d}x \sqrt{-g}\left(R-2\Lambda+\sum_{k=2}^K n_k c_k E^{2k}\right) \label{action}
\ee
with $\lp$ the $d$ dimensional Planck length and where $E^{2k}$ is the $2k$-dimensional Euler density, which is topological in $d=2k$ and zero for $d<2k$. The normalization $n_k$ is a number such that with $\Lambda=-(d-1)(d-2)/L^2$ an $AdS$ vacuum exists with radius $L^2/\ff$, with $\ff$ satisfying
\be
\Upsilon[\ff]\equiv \sum_{k=0}^K (-1)^k c_k \ff^{k}=0.\label{adsvacua}
\ee
For definiteness it might be useful to keep in mind the Gauss-Bonnet lagrangian, which is non-topological for $d\geq 5$ and has been extensively studied in the literature. In this case we have 
\be
I_{GB}=\frac{1}{\lp^{d-2}}\int d^d x\sqrt{-g}\left(R+\frac{(d-1)(d-2)}{L^2}+\frac{\lambda L^2}{(d-3)(d-4)} E_4\right)
\ee
with $E_4=R_{abcd}R^{abcd}-4 R_{ab}R^{ab}+R^2$. The $AdS$ vacua are determined by $1-\ff+\lambda \ff^2=0$.

We will consider that the action \reef{action} is coupled to an arbitrary matter sector,
\be
I_M=\int d^d x\sqrt{-g} \mathcal L_M,
\ee
and the matter stress tensor is defined as
\be
T_{ab}=-\frac{2}{\sqrt{-g}}\frac{\delta S_M}{\delta g^{ab}}.
\ee
We will not permit curvature couplings to the matter lagrangian in this work. This implies for instance that, if covariant derivatives appear in $L_M$, they should be symmetrized. 

\subsection{Black hole solutions}
In the following we will be interested in a class of backgrounds describing black hole solutions. Early work on such solutions in Lovelock theories of gravity can be found in references \cite{Cai:1998vy,Cai:2001dz,Cai:2003kt}.
We take a metric of the form:
\be
ds^2=-\left(\kappa+\frac{r^2}{L^2}f(r)\right) \frac{dt^2}{\ff}+\frac{L^2 dr^2}{\kappa+\frac{r^2}{L^2}\,g(r)}+r^2 (\ud \Sigma^{d-2}_\kappa)^2, \label{bgmetric}
\ee
with $\ud \Sigma$ is the volume element for the space with unit radius and constant 
curvature\footnote{More concretely, for $\kappa=1$ we have $d\Sigma=dS_{d-2}$, the volume element for a $(d-2)$ sphere; for $\kappa=-1$ $d\Sigma=dH_{d-2}$ the volume element for the $d-2$ hyperbolic space. Finally for $\kappa=0$ we take $d\Sigma^2=L^{-2}\sum_i (dx^i)^2$.}
 $\kappa=-1,0,1$.  In the above the factor $\ff$ is chosen to obtain a unit speed of light at the boundary. The functions $g(r)$ and $f(r)$ satisfy $g(r_0)=f(r_0)=-L^2\kappa/r_0^2$, so that there is a horizon at $r=r_0$. The metric is otherwise regular and smooth. 

The entropy of these black holes is computed by means of the Wald formula   \cite{Wald:1993nt,Iyer:1994ys} (see also \cite{Jacobson:1993vj}):
\be
S_{BH}=-2\pi \oint \sqrt{h}\, \frac{\delta \mathcal L}{\delta R_{abcd}}\, \epsilon_{ab}\epsilon_{cd} \label{WaldS}
\ee
where the integral is taken over the black hole horizon, $\mathcal L$ is the gravitational lagrangian and $\epsilon_{ab}$ are the horizon binormals.
Using the results of appendix \reef{waldform} we find
\be
S=4\pi V_{d-2}\left(\frac{r_0}{\lp}\right)^{d-2}  \sum_{k=1}^K \frac{(d-2)k}{d-2k}\, c_k  \left(\kappa\frac{L^2}{r_0^2}\right)^{k-1} \label{bhentr},
\ee
where $V_{d-2}$ will generically stand for the volume of the transverse space, regardless of the value of $\kappa$. In the particular case where there is no matter present, exact black hole solutions can be found, with $f(r)=g(r)$ and
\be
\Upsilon[g(r)]\equiv \sum_{k=0}^K c_k g(r)^{k}=\left(\frac{r_0}{r}\right)^{d-1}\Upsilon[g(r_0)].
\ee
While this algebraic equation has many solution branches, there is only one (the one corresponding to the smaller positive root) which is ghost-free and regular (see \cite{Boulware:1985wk} and also \cite{Camanho:2009hu,deBoer:2009pn} for more recent papers analysing some of these issues).

\section{Holographic c-functions and beyond}
\label{sec3}
In the presence of a matter sector, there could be several stable $AdS$ vacua with different radius of curvature. It is then possible to consider backgrounds which interpolate between such two vacua, which would describe a renormalization group flow between two conformal field theories
with different central charges. Such ``RG flow'' backgrounds take the generic form
\be
ds^2=d\beta^2+e^{2 A(\beta)}\left(-\ud t^2+\ud \mbf x^2\right). \label{rgbg}
\ee
With $A(\beta)=\sqrt{\ff} \beta/L$ the above describes an $AdS$ background with radius $L/\sqrt{\ff}$. In general $A(\beta)$ will be some function interpolating between two such $AdS$ spaces located at $\beta=\pm \infty$.

In Einstein gravity it is possible to define a holographic $c$-function \cite{Freedman:1999gp}. This takes the simple form
\be
c(\beta)=\left(\frac{1}{\lp}\right)^{d-2} \frac{1}{A'^{d-2}}.
\ee
By construction, when evaluated on $AdS$ spaces this function gives the ratio of the $AdS$ length to the Planck length to the power of $d-2$, which is exactly as required of a $c$-function, since in Einstein gravity, all conformal anomaly coefficients are proportional to this quantity. The second requirement of a $c$-function is that it is monotonous along RG flows. This follows directly from the Einstein equations for such a background which give
\be
(d-1)A''(\beta)=\lp^{d-2}\left(T^{t}_{\ t}-T^{\beta}_{\ \beta}\right) \label{EinsEqn}.
\ee
If we demand that the null energy condition \cite{LargeScale} is satisfied, the right hand side of the equation above is negative. This immediately implies  that $c(\beta)$ monotonously decreases as one moves towards the interior of the geometry.

We would now like to generalize this construction to Lovelock theories of gravity. Our strategy will be to find the equations of motion for the background and from that guess what the correct $c$-function should be. We shall however be more general than up to now, by considering black hole metrics such as \reef{bgmetric}. Notice that the RG flow geometries \reef{rgbg} may be obtained from these by choosing $\kappa=0$ and performing the change of variables
\be
r\to L e^{A(\beta)}, \qquad g(r)\to (L \, A'(\beta))^2. \label{varchange}
\ee
The motivation for this generalization is that such geometries are still easily tractable, and should also describe renormalization group flows of a dual field theory, albeit in a non-trivial state. We will see that the equations of motion take a very simple form for these backgrounds, which ultimately justifies our interest in studying them.

\subsection{Lovelock theory: equations of motion}

Given the total action $S=S_{g}+S_M$ we will be interested in gravitational backgrounds which are static and preserve translational invariance along spatial directions. The most general metric ansatz with these characteristics is the familiar black hole solution given in \reef{bgmetric}.
For Lovelock gravity, the equations of motion describing this background are easily found, as shown in appendix \reef{Lovelock}. Consider first the $tt$ component of the Einstein equations,
\be
E_t^t=\frac{\lp^{d-2}}{2} T_t^t\equiv -\frac{\lp^{d-2}}{2} \rho,
\ee
where we have defined the energy density $\rho$. The sign is such that a free scalar field will have positive $\rho$. In more detail, the equation takes the form
\be
\frac{\ud}{\ud r} \left(r^{d-1}\Upsilon[g]\right)=\frac{L^2\, \lp^{d-2}}{d-2} \,\,r^{d-2}\rho
\ee
with
\be
\Upsilon[g]=\sum_{k=0}^K c_k g^k(r)=1-g(r)+\lambda g(r)^2+\ldots
\ee
The radial equation $E_r^r$ also takes a simple form. We obtain
\be
(d-1) \Upsilon[g]+\Upsilon'[g]\, 
\left(\frac{\kappa\, L^2+r^2\,g }{\kappa\, L^2+r^2\,f}\, r f'
+\kappa \frac{f-g}{\kappa\, L^2+r^2\,f}\right)
=\frac{L^2\, \lp^{d-2}}{d-2}T_r^r\equiv \frac{L^2\, \lp^{d-2}}{d-2} p_r \label{radeq}
\ee
where we have defined the ``radial pressure'' $p_r$. In particular notice that
\bea
E_t^t-E_r^r&=&\frac{d-2}{2}
\frac{(-\Upsilon'[g])}{L^2+r^2 f}\left[\frac {\kappa L^2} r\left(r^2 g-r^2 f\right)'+r^3 (f g'-g f')
\right]
\\
&=&-\frac{\lp^{d-2}}{2}\left(\rho+p_r\right)\label{neceq}
\eea
If the null energy condition is satisfied the righthand side of this equation is negative. We will come back to this point later on. Going back to the $tt$ equation, we can integrate to find
\be
\Upsilon[g]=\frac{L^d\,\lp^{d-2}}{d-2} \frac{\int_{r_0}^r \ud r' \, (r'/L)^{d-2} \rho(r')}{r^{d-1}}\equiv
\frac{L^d}{(d-2)V_{d-2}}
\lp^{d-2}\frac{M(r)}{r^{d-1}}. \label{alg}
\ee
Here we have defined $M(r)$ as a measure of the energy contained in the spacetime from some $r_0$ up to $r$. For instance, if a black hole horizon is present then $r_0$ would be its location, and one would have\footnote{In this case the density $\rho$ has a $\delta$ function contribution at the horizon. This is because we are working in coordinates appropriate to Schwarzschild observers which have no access to information beyond the horizon at $r_0$.} 
$$
M(r)=(d-2)\frac{V_{d-2}}{\lp^{d-2}} \frac{r_0^{d-1}}{L^d}\Upsilon[g(r_0)]+\ldots
$$ where the dots represent additional matter contributions.

We could now solve algebraically for $g$ in \reef{alg} and use the $rr$ equation to determine $f$. This would completely determine the spacetime geometry if $\rho$ and $p_r$ were known. However, it is instructive to go back to the $tt$ equation. Using \reef{alg} we obtain
\be
\left(-\Upsilon'[g]\right ) \frac{\ud g}{\ud r}=\frac{2 L^d}{d-2}\,  \frac{\ud \Psi}{\ud r}. \label{graveq}
\ee
where we have defined the ``gravitational potential'':
$$\Psi\equiv -\frac{\lp^{d-2}}{2 V_{d-2}}\frac{M(r)}{r^{d-1}}.$$
The expression \reef{graveq} has a simple interpretation. The function $g$ here plays a role analogous to the red-shift: we shall call it the {\em rad-shift}, for radial-coordinate. Since $-\Upsilon'[g]$ is always positive,
%
\footnote{To see why $-\Upsilon'[g]$ must be positive, one simply notes that, if $M(r)$ is positive and non-decreasing, then the solution to the algebraic equation \reef{alg} develops a singularity at some $r$ unless the polynomial $\Upsilon[g]$ is monotonously decreasing up to its first root.}
%
the above tells us that the rad-shift increases towards the $AdS$ boundary if the ``gravitational field'' in the radial direction, $-\Psi'(r)$ is pointing towards the IR. In other words, the gravitational field tells us about the direction in which the rad-shift $g$ decreases. 


A particular case of the equations developed above is given by $f(r)=\ff$, $\kappa=0$. Performing the change of variables given in  \reef{varchange} we go back to the RG flow background \reef{rgbg}. However, generically now one expects the gravitational field to be pointing towards the UV, not the IR. This is because, for $A(\beta)=\sqrt{\ff} \beta/L$, the RG flow metric becomes that of an $AdS$ geometry with radius $L/\sqrt{\ff}$. Under an RG flow we expect that the radius of the $AdS$ space shrinks from the UV to the IR, which correlates with a shrinking of the central charge of the dual field theory. This requires $A''(\beta)\leq 0$, and therefore $g'(r)\leq 0$. This is exactly opposite to our expectations for the rad-shift in a black hole background, where $g$ should decrease toward the interior of the geometry.

The answer to this apparent puzzle is that since the gravitational field is now pointing towards the UV, it must be that the matter density $\rho$ is negative, contrary to what happens in the presence of a ``normal'' matter source. To see exactly how this happens, consider the radial equation \reef{radeq} which now becomes simply
\be
p_r=-\frac{M(r)/V_{d-2}}{r^{d-1}}.
\ee
In RG flow backgrounds we typically demand that the null energy condition is satisfied, which in this case amounts to $\rho+p_r\geq 0$. From the above it is clear however that such a condition requires that the matter energy density $\rho$ is generically negative. This is because in the asymptotic UV $AdS$ space one has $\rho=0$ but non-zero $M(r)$. Therefore the null-energy condition necessarily gives $M(r)\leq 0$ which requires $\rho\leq 0$ in a large region of the geometry.

\subsection{The $\mathcal N(r)$ flow equation}

The equation \reef{graveq} plays a role analogous to equation \reef{EinsEqn} in the Einstein case. As such we need to find an expression whose radial derivative is proportional to the lefthand side of \reef{graveq}. This expression, which we shall call the $\mathcal N$-function is given by:
\be
\mathcal N(r)=\frac{1}{g^{\frac{d-2}2}}\left(\sum_{k=1}^K \frac{(d-2) k}{d-2k}\,c_k (-g)^{k-1}\right).\label{Ndef}
\ee
Taking its derivative and using \reef{graveq} we find
%
%
\be
\frac{\ud \mathcal N}{\ud r}= \left(\frac{L}{\sqrt{g}}\right)^{d}\left(- \frac{\ud \Psi}{\ud r}\right) \label{Ngrav}
\ee
This is one of our main results. It tells us how the function $\mathcal N(r)$ varies depending on the local ``gravitational field'', $-\Psi'$. This is not a trivial result, i.e. a repackaging of the equations into some complicated function $\mathcal N$. This is because as we shall shortly see, this function defines a $c$-function on RG flow backgrounds, and it is also related to black hole entropy. It is thus naturally interpreted as counting degrees of freedom. From the above it is clear that $\mathcal N(r)$ doesn't have a well defined monotonicity, since as we've seen the sign of the gravitational field depends on whether there is positive or negative mass. As such, while $\mathcal N(r)$ is separately monotonous in vacuum black hole solutions or in RG flow backgrounds containing matter satisfying the null energy condition, it is not monotonous say when one starts off on an $AdS$ geometry in the UV and flow down to an IR $AdS$ geometry containing a black hole. 

Applying this equation to the case where $\kappa=0$, $f(r)=\ff$ and we get back to the RG flow background described by metric \reef{rgbg}. 
In this case we have
\be
\mathcal N= \frac{1}{(L\,A')^{d-2}}\left(\sum_{k=1}^K (-1)^{k-1}\frac{(d-2) k}{d-2k}\,c_k (LA')^{2(k-1)}\right).
\ee
When the mass density is constant, such as when there is only a cosmological constant or a fixed scalar at the bottom of a potential, the gravitational field is zero and there is no flow of $\mathcal N$. More generally, as we've shown in the previous section, the null energy condition implies that the mass $M(r)$ is negative, and therefore $\mathcal N(r)$ is monotonously increasing from the IR to the UV.
At the fixed points we obtain
\be
\mathcal N= \frac{1}{\ff^{(d-2)/2}}\left(\sum_{k=1}^K \frac{(d-2) k}{d-2k}\,c_k (-\ff)^{k-1}\right)\equiv \left(\frac{\lp}{L}\right)^{d-2} A \label{NEuler}.
\ee
As we show in appendix \reef{Euler}, the coefficient $A$ defined in the above turns out to be precisely given by
the Euler anomaly for Lovelock theory. Of course, this definition only makes sense for odd-dimensional gravity theories. In general, $A$ is expected to be related to a universal coefficient appearing in entanglement entropy calculations of the dual field theory \cite{Myers:2010xs,Myers:2010tj}. We see then that Lovelock theories support a $c$-function $A(\beta)$, defined by 
$$
A(\beta)=\left(\frac{L}{\lp}\right)^{d-2}\mathcal N(\beta)
$$
This supports the idea that $\mathcal N$ characterizes the number of degrees of freedom along RG flows. 

\subsubsection{The black hole case}

Let us consider the case where we have a black hole present, and no matter. In this case the gravitational potential becomes
\be
\Psi=-\lp^{d-2} \left(\frac{r_0}{r}\right)^{d-1}\Upsilon[g(r_0)],
\ee
with $g(r_0)=-\kappa L^2/r_0^2$. The flow equation \reef{Ngrav} immediately tells us that $\mathcal N(r)$ is monotonously {\em increasing} as the flow proceeds from the boundary towards the horizon. There are now three distinct possibilities, according to the value of $\kappa$. 

\begin{itemize}
\item $\mathbf {\kappa=-1}$ \\ In this case $g(r)$ is always positive. Then it is easy to see that $\mathcal N(r)$ monotonously increases from the boundary (where it is related to the $A$ coefficient by equation \reef{NEuler}) to the horizon, where it turns out to be related to the black hole entropy, as given in \reef{bhentr}:
\be
S_{\mbox{{\tiny BH}}}=4\pi V_{d-2}  \left(\frac{L}{\lp}\right)^{d-2}\mathcal N(r_0)
\ee
In this way, for such geometries the $\mathcal N$ function nicely interpolates between the $A$ anomaly in the UV and the black hole entropy in the IR.
\item $\mathbf {\kappa=0}$ \\ In this case $g(r)$ has a zero at $r=r_0$, the horizon. The function $\mathcal N(r)$ is still monotonously increasing from the boundary to the horizon, but it diverges there. To regulate this divergence we evaluate $\nn$ not at the horizon but at some finite distance. In analogy with the $\kappa\neq 0$ cases, we take $g=\frac{\lp^2}{r_0^2}$. With this regularization we also get a relation between $\mathcal N$ at the horizon and the entropy, namely
\be
S_{\mbox{{\tiny BH}}}=4\pi V_{d-2}\mathcal N(r_0)
\ee
where the integral is taken at the horizon.
The divergence implies there is a dramatic increase from an order one to order $(L/\lp)^{d-2}$ in the number of effective field theory degrees of freedom as one approaches the black hole horizon. 

\item $\mathbf {\kappa=1}$ \\ This case is the most intriguing. The function $\mathcal N$ diverges at $g=0$, but this does not correspond to the position of the black hole horizon, which occurs at $g=-L^2/r_0^2$, a negative value for $g$. In fact, our expressions \reef{Ndef},\reef{Ngrav} can even become imaginary or negative there, depending on $d$. Nevertheless, we can find that the entropy of the black hole is now given by
\be
S_{\mbox{{\tiny BH}}}=4\pi V_{d-2} \left(\frac{L}{\lp}\right)^{d-2}  |\mathcal N(r_0)|. 
\ee
In particular $\mathcal N$ is perfectly finite at the horizon. Why the $\mathcal N$ function should present a divergence at $g=0$ is left to be understood, although we will present some possibilities in the discussion. 

\end{itemize}

\section{The effective phase space}
\label{sec4}
The results of the previous section indicate that $\nn$ describes some kind of effective number of degrees of freedom as a function of scale. At black hole horizons, it is directly proportional to the entropy, with a well known proportionality factor. Since the entropy in higher derivative theories is given by the Wald formula \reef{WaldS}, this hints that we might be able to write $\nn$ in terms of it.  This is further suggested by the fact that $\mathcal N$ is related to the Euler anomaly on $AdS$ spaces, and this is also given by the Wald formula, as shown in appendix \reef{Euler}. 

Consider a radial foliation of the background \reef{bgmetric}. More concretely, choosing two vectors $n^a$, $m^a$ with non-zero components
\be
n_r=\sqrt{g_{rr}}, \qquad m_t=\sqrt{-g_{tt}},
\ee
we take
\be
h_{ab}=g_{ab}-n_a n_b+m_a m_b,
\ee
and $\epsilon_{ab}=2 n_{[ a} m_{b]}$. Notice that $h_{ab}$ only has non-zero components when $a,b$ run through the $(d-2)$ transverse directions, denoted by indices $i,j,k,\ldots$. It defines the metric of the constant radial surfaces that constitute the foliation.
In appendix \reef{Lovelock} we show that the Wald formula applied on these surfaces gives%
\be
S= -2\pi V_{d-2}\sqrt{h} \frac{\partial \mathcal L}{\partial R_{abcd}}\epsilon_{ab}\epsilon_{cd}
=4\pi\, V_{d-2}\,\left(\frac{L}{\lp}\right)^{d-2}\sum_{k=1}^K \frac{(d-2)k}{d-2k}c_k (-g)^{k-1},
\ee
and therefore we can write:
\be
\nn=\frac{S}{4\pi \Omega_{\mbox{\tiny{eff}}}}, \label{NSO}
\ee
This expression constitutes one of our main results. It describes the $\mathcal N$-function as the ratio of an entropy to an interesting quantity,  
\be
\oef=\left(\frac{L}{\lp}\right)^{d-2}\! \left(\frac{r}{L^2}\, g(r)\right)^{d-2} V_{d-2}. \label{oef}
\ee
We shall argue that this represents a kind of effective phase space volume. Let us now examine in more detail the two ingredients going into \reef{NSO}.

\subsection{Entropy and information content}

The holographic principle states that gravitational dynamics in a closed region can be equivalently described by the dynamics on its bounding surface. Let us imagine that there is a holographic screen enclosing such a region. If nothing is known about what is going on behind the screen, we can associate to it an entropy characterizing our ignorance, In other words, a given region has an associated information content, which is the logarithm of the total number of states which can be described in such a region. In a normal field theory this entropy should be proportional to the volume, but in gravitational theories this is not so because of the possibility of black hole formation \cite{'tHooft:1993gx}. When a black hole completely fills a certain region of spacetime, we have maximal ignorance about that region, and so the information content of a region should be related to the entropy of a hypothetical black hole which would completely fill that region. Since the Bekenstein-Hawking entropy of a black hole is simply the ratio of its area to Newton's constant, this directly leads to the proposal that the information content of a region scales like the area, and hence the notion of holography.

For more complicated gravitational theories, the entropy of a black hole is no longer simply given by the area, but rather by the Wald formula \reef{WaldS}
Following the same logic as in \cite{'tHooft:1993gx}, one is led to the idea that the information content of a given region of space time is given by Wald's formula, but now the integral should be taken over any closed surface, not necessarily a horizon. It is in this information sense that the entropy $S$ appears in the formula \reef{NSO} for the $\mathcal N$-function, and not as a genuine entropy: at no point are we actually coarse graining or integrating out the geometry behind the screen. In the following, by abuse of language, we will refer to the (log of the) number of states as an entropy, even though no such tracing over state has been necessarily made.


\subsection{The effective phase space}

Now consider the factor $\oef$. In a conformal relativistic field theory, the entropy of a thermal state at a given temperature can be thought of as the product of the number of available degrees of freedom, $\mathcal N$, by the phase space volume $\Omega$ available to each one of them. For instance, the entropy of a hot, large $N$, non-abelian plasma in $D$ dimensions is of the form
\be
S= C\times N^2 V_{D-1} T^{D-1}\simeq \mathcal N \times \Omega
\ee
with $\mathcal N\simeq C$ and $\Omega\simeq N^2 V_{D-1} T^{D-1}$. Notice we have included the different gauge polarizations in the definition of phase space. This means that $\mathcal N$ here counts the number of fields, these fields consisting themselves of several components which tranform into each other under global gauge transformations. The temperature  acts a soft cut-off on the allowed energy and momentum states, since for energies higher than $\simeq T$ the probability of finding such a state decays exponentially. We expect that the logarithm of the total number of states, with IR and UV cut-offs, that can be constructed in a conformal field theory should once again be of this form, but with $T$ replaced by $\Lambda$, the momentum space cutoff.

With these considerations, the number of degrees of freedom $\mathcal N$ available at some fixed cut-off $\Lambda$ is given by
\be
\mathcal N\simeq \frac{S}{\Omega}\bigg |_{\Lambda}.
\ee
Then it appears that $\oef$ should be playing the role of an effective phase space at a given cut-off, as defined by a fixed radial coordinate. In fact, inverting this logic, we should be able to extract the cut-off from the expression for phase space. For the RG flow backgrounds \reef{rgbg}, we have
\be
\oef=\left(\frac{L}{\lp}\right)^{d-2} \times V_{d-2}\times  \left(e^{A} A'\right)^{d-2}
\ee
This takes the form of a ``gauge volume'', $(L/\lp)^{d-2}$, times a real space volume $V_{d-2}$, which alternatively acts an IR cut-off; from the remaining factor we can identity as the momentum space cut-off:
\be
\Lambda_{\mbox{\tiny eff}}=e^{A} A'.
\ee
It is interesting to notice that in \cite{Heemskerk:2010hk} it was argued that in these backgrounds the field theory cut-off should be roughly given by this expression. If our interpretation of $\oef$ as a phase space is correct, this would make this statement more precise.

For the general case, if we compute $K_{ab}=\nabla_a n_b$ we get
\be
K_{ij}=-\frac{h_{ij}}L \sqrt{\frac{L^2}{r^2}\kappa+\, g(r)}.
\ee
Then it follows that
\bea
\oef &=& \left(\frac{L}{\lp}\right)^{d-2}\oint d^{d-2}x \sqrt{h}\,\sqrt{\mbox{det}\left( K_i^k K_k^j-\,\kappa \,\delta_i^j \frac{L^2}{r^2}\right)} \label{kkeqn}
\eea
where the determinant should be taken over the $(d-2)$ transverse directions where $h_{ij}\neq 0$. Notice that the above has the structure $\simeq \Lambda^2-m^2$, precisely as expected if we are to interpret $\oef$ as counting states. This strongly suggests that the dual field theory cut-off scale is directly related with the extrinsic curvature $K_{ab}$ of a constant radial surface, as defined by the vector $n_a$. In the particular set of geometries we are considering we can take:
\be
\Lambda^{d-2}=\sqrt{h}\,\mbox{det}\left(K_i^j\right).
\ee
In this way, satisfyingly, the field theory momentum is directly associated with $K_{ij}$, which is the canonical momentum of the metric, associated with radial translations.

Notice there's an alternative way of writing \reef{kkeqn}, using
\be
K_i^k K_k^j-\,\kappa \,\delta_i^j \frac{L^2}{r^2}=R_{ik}^{\ \ jk},
\ee
where the implicit sum over $k$ runs only over the transverse directions. In this way we have connected the effective phase space volume with the Riemann curvature components in the transverse space. 

\subsection{Effective phase space and gravity as an entropic force}

The flow equation \reef{Ngrav} relates the variation of $\mathcal N$ with the gravitational field. At the same time, from \reef{NSO} we can relate $\mathcal N$ to the ratio of the entropy to a quantity we have interpreted as a phase space. From these equations we can write
\be
\frac{\ud S}{\oef}=\left(\frac{L}{\sqrt{g}}\right)^d \ud \Psi+ \mathcal N\, \ud \log(\oef)\label{varS1}
\ee
This is reminiscent of an equation found by Verlinde \cite{Verlinde:2010hp}, relating the ``depletion of energy per bit'' to the Newtonian potential (equation (3.16) in that reference). In this language it would be more natural to work with $\Upsilon[g]$ instead of $\Psi$, using
\be
\Upsilon[g]=-\frac{2}{d-2} L^d \Psi. \label{ups}
\ee
The quantity $\Upsilon[g]$ is zero at the $AdS$ boundary, and in the planar black hole case, equals one at the black hole horizon, thus providing the coarse-graining variable mentioned in \cite{Verlinde:2010hp}. There are however two important differences with the above equation. The first is that we interpret $\oef$ as a phase space volume and not as a number of degrees of freedom. Secondly there is an extra term in the above because the change in entropy is coming from moving along the geometry, and not by adding extra matter to it as in \cite{Verlinde:2010hp}. One way of directly deriving Verlinde's formula is as follows. We consider varying the matter content in a given region, keeping the phase space $\oef$ fixed. This relates the variations in the rad-shift $g$ and $r$ via $d(r^2 g)=0$. This makes sense: if the holographic screen was precisely located at a black hole horizon, adding a small mass would shift the position of the horizon swallowing the screen, so we must shift its position at the same time. The condition above guarantees that the screen would follow the horizon. The rad-shift $g$ is fixed by the previous equation that relates $\Upsilon[g]$ with the gravitational potential. By taking the variation of \reef{ups}, and of the entropy formulae \reef{s1},\reef{llwald} subject to the constraint $d(r^2 g)=0$, we indeed precisely find
\be
\frac{\ud S}{\oef}=\left(\frac{L}{\sqrt{g}}\right)^d \ud\Psi.
\ee
Although expected, this does not trivially follow from \reef{varS1} - in that equation the variation is with respect to $r$ keeping the mass fixed, whereas in the present case we vary the mass and $r$ keeping $\oef$ fixed. As such, this is yet another indication that $\oef$ plays an important role in the holographic interpretation of the equations, and in particular in the possible interpretation of gravity as an entropic force.

\section{More on holographic phase space}
\label{sec5}

In this section we shall take a new tack on an old proposal \cite{Susskind:1998dq}, that the number of degrees of freedom holographically associated to a given surface is simply proportional to the area of that surface divided by an appropriate power of the Planck length. Here we shall also use the area of a surface as a basic ingredient, but we will interpret the result differently - we propose that the the proper length divided by Planck length should actually be interpreted as a phase space volume - a number of states, not of degrees of freedom. This should be taken as an alternative proposal for a gravitational expression for the phase space volume, different from the $\oef$ we have defined previously. We shall see that the present definition leads to some suggestive results. In section \ref{thermoent} we will have more to say on the connection between these two notions of phase space.

Let us then define the phase space volume associated with a region $M$ as the area of its boundary,
\be
\Omega=\oint_{\partial M} \frac{dA}{\lp^{d-2}},\label{defOm}
\ee
and as an example, focus on a specific geometry, namely the vacuum $AdS$ solution described by the case $\kappa=0$, $g=\ff$ of the metrics \reef{bgmetric}.
Applying the above definition, with $\partial M$ a constant $r,t$, surface gives:
\be
\Omega=\left(\frac{L}{\lp}\right)^{d-2} V_{d-2} \left(\frac{r}{L^2}\right)^{d-2}. \label{omx}
\ee
In this expression we have the product of a volume $V_{d-2}$, on which the dual field theory lives, by something with units of momentum, $r/L^2$. It follows that the interpretation of $\Omega$ with a phase space is consistent if we take $r/L^2$ as the dual field theory momentum cut-off. The remaining factor of $L/\lp$ accounts for the fact that phase space also includes gauge-group polarizations. For instance in $N=4$ $SU(N)$ super Yang-Mills theory, $(L/\lp)^3\simeq N^2$.

Let us go back to the more general geometry, and define, in analogy with the previous section, a number of degrees of freedom $N_{\mbox{\tiny{dof}}}$, by taking
\bea
N_{\mbox{\tiny{dof}}}\equiv \frac{S}{4\pi \Omega} = 2 \frac{\partial \mathcal L}{\partial R_{rt}^{\ \ rt}}.
\eea
This is precisely the expression for the ``surface density of degrees of freedom'' which has appeared in \cite{Padmanabhan:2010xh}. This further supports our interpretation of $\Omega$ as a phase space. In the backgrounds \reef{bgmetric} this reduces to
\be
N_{\mbox{\tiny{dof}}}=\sum_k \frac{(d-2)k}{d-2k}c_k g^k. \label{ndof}
\ee
It is apparent that $N_{\mbox{\tiny{dof}}}\neq \mathcal N$. The extra factors of $g$ in the denominator appearing in the definition of $\mathcal N$ are crucial in order to obtain the flow equation \reef{Ngrav}; no such equation is satisfied by $N_{\mbox{\tiny{dof}}}$.

As it stands, it is not immediately clear that this interpretation of $\Omega$ as a phase space holds water. After all, the expression \reef{omx} is always the same, regardless of whether one is considering a background with or without a black hole and different transverse space. But, we expect that the phase space structure to be quite different for these. For instance, in the global $AdS$ there is a mass gap, and accordingly the available phase space should no longer scale as the momentum. To see the differences between these geometries we need to change coordinates. Since  $r$ has no immediate geometrical meaning, it is  better to work with the proper radial distance, 
\be
\beta=\int_{r_0}^{r} \ud r' \sqrt{g_{rr}},
\ee
where $r_0$ is some IR cut-off. It should be taken as the place where the geometry ends, which indeed for a black hole metric would be at the horizon $r=r_0$. By definition $\beta=0$ at this point. Also it is easy to see for asymptotically large $r$ we have
\be
\beta\simeq L\log{r/r_0}.
\ee
In empty $AdS$, $\beta$ is the logarithm of the dual field theory energy scale, and is the natural quantity that parameterizes the RG flow.  In the following we shall examine the behaviour of $\Omega$ with $\beta$ for a few different geometries. We will start by considering first the planar and then the spherical horizon $AdS$-Schwarzschild black hole solutions. We also stick to five-dimensions and Einstein gravity in what follows, though the lessons we shall derive should apply to more general cases.

\subsection{$AdS$-Schwarzschild black hole}
The $AdS$-Schwarzschild black hole solution of Einstein gravity with a negative cosmological constant is
\be
ds^2=\frac{L^2\,dr^2}{r^2 g(r)}+\frac {r^2}{L^2}\left(-g(r)dt^2+d\mbf x^2\right)
\ee
with $g(r)=1-(r_0/r)^4$. There is a horizon located at $r=r_0$, and the temperature is $T=r_0/\pi L^2$. 
In terms of the proper distance $\beta$ we have
\be
r=r_0\sqrt{\cosh(2\beta/L)}.
\ee
According to our definition, the phase space volume corresponding to a given direction, say $x$ is given by
\be
\Omega_x=\frac{L}{\lp}\times R\times \pi T \times \sqrt{\cosh(2\beta)} 
\ee
This is an intriguing result. If it weren't for the square root, this would be (up to a constant prefactor) the canonical partition function of a fermionic oscillator, or a two state system at inverse temperature $\beta$. The square root indicates that we have here only half of a fermion. Actually, we can rewrite:
\be
\sqrt{\cosh(2\beta)}=\sqrt{2\cosh(\beta)^2-1}
\ee
and in this form this precisely matches the partition function of an anyon harmonic oscillator \cite{BoschiFilho:1994an}. This is suggestive of an equivalence between a classical microcanonical partition function, or phase space volume at a given cut-off, and a canonical partition function at an inverse temperature $\beta$ related to this cut-off.

Thinking of the two state system, or spin,  we see that as we approach the black hole horizon the $\beta$ parameter goes to zero, i.e. the ``temperature'' goes to infinity and the spin becomes completely randomized. Pushing this analogy further, let us compute the ``mean energy'':
\be
\langle E \rangle=-\frac{d \log \Omega}{d \beta}=-\sqrt{g_{rr}}\, \frac{d \log(\sqrt{h})}{dr}=h^{ab}K_{ab} \label{EKeqn}
\ee
with $K_{ab}$ the extrinsic curvature associated to the $r$ foliation defined by the vector $n=(0,\sqrt{g_{rr}},0,0,0)$.
Remarkably the mean energy is given by a natural geometric observable. Similarly, the mean energy squared is
\be
\langle E^2 \rangle =\frac{1}{\Omega}\frac{d^2\Omega}{d \beta^2}=-R_{abcd}n^{a}n^{c}h^{bd}.
\ee
So at least some of the curvature components have a ``thermodynamical'' interpretation, if we identify $\beta$ as an inverse temperature.
Finally, notice that we can write $\langle E\rangle$ as the sum of three separate contributions
\be
\langle E\rangle=-\sum_{i=1}^{d-2}\frac{d\log \Omega_i}{d\beta}\equiv -\sum_{i=1}^{d-2}\mathcal E_i
\ee
which can be understood as a sum of the average energies corresponding to each of the $(d-2)$, in this case three, transverse directions. The quantity $\mathcal E_i$, defined by $d\log \Omega_i/d\beta$ has an interesting meaning. It is telling us by how much the logarithm phase space volume is changing when we change the RG parameter $\beta$. For the current geometry this gives
\be
\mathcal E_i=\frac{\sqrt{g}}{L}.\label{Ei}
\ee
In the asymptotic $AdS$ region, this simply a constant. This means that $\beta$ and $\log \Omega$ are essentially the same. In this region, the RG flow coincides with the flow in scale. However, as we approach the black hole horizon, this is modified dramatically. There, although the $\beta$ parameter is going to zero, the phase space volume $\Omega$ is becoming a constant. This is intriguing, because at the horizon scale there is another quantity which is vanishing: because of the deconfined nature of the dual field theory plasma, quantum correlations vanish for scalar larger than $1/T$. This suggests that $\beta$ might have more to do with correlations than with scale. We will offer some speculations on this in section \ref{thermoent}.

\subsection{Global $AdS$ black hole}

Now let us consider global $AdS$-black holes in Einstein gravity. These are described by the metric
\be
ds^2=\frac{\ud r^2}{1+\frac{r^2}{L^2} \,g(r)}-\frac{r^2}{L^2}\left(1+\frac{r^2}{L^2}\,g(r)\right) \ud t^2+\frac{r^2}{L^2}\,\ud S_3^2 \label{globalBH}
\ee
with $\ud S_3$ is the volume element on the three sphere, $\ud S_3=\sin^2(\theta)\sin(\psi) \ud \theta\, \ud \psi \,\ud \phi$.
When $g(r)=1$ we have $AdS$ in global coordinates. In this case the phase space expression is given by
\be
\Omega=\left(\frac{L}{\lp}\right)^3\int \ud S^3=\left(\frac{L}{\lp}\right)^3 V_{S_3} \left(\frac{r}{L}\right)^3.
\ee
One might think that the interpretation of the $r/L$ factor is the angular momentum scale, in analogy with the previous case. This however cannot be true, since we know there is a mass gap in the dual theory. Accordingly the number of states must go to zero at the mass gap scale. This is exactly borne out by our expression. To see this first notice that in terms of $\beta$ we now have
\be
\frac{r}{L}=\sinh\left(\frac{\beta}{L}\right)=\frac{\sqrt{\cosh \left(\frac{2\beta}{L}\right)-1}}{\sqrt{2}}
\ee
Interestingly we now get something which resembles the partition function of a fermionic harmonic oscillator.
For large $\beta$ the above reduces to $\beta\simeq L \log(r/L)$. However, for small $\beta$ the extra term under the square root becomes important. The above should be interpreted as
\be
\frac{r}{L}=\frac{L}2\sqrt{\Lambda^2-m^2}
\ee
where $\Lambda$ is the energy scale $L^{-1}\sqrt{2\cosh \left(\frac{2\beta}{L}\right)}$, and $m=\sqrt{2}/L$ is the mass gap. Our expression for the phase space, $r/L$, counts the number of states, as it should. Also, the relation between the energy scale and the $\beta$ parameter is non-trivial. Indeed it is the same as in the Poincar\'e patch black hole replacing $r_0$ with $L$. This is because, analogously to before, the correlations go to zero as one approaches the scale of the mass gap.
The expression for the energy scale can also be written in terms of $r$ as
\be
(L \Lambda)^2=1+2\frac{r^2}{L^2}
\ee
When there's a black hole present, we have $g(r)=1-\frac{r_0^4}{r^4}\left(1+\frac{L^2}{r_0^2}\right)$, and it is easy to check that there is a horizon at $r=r_0$. The temperature is given by
\be
T=\frac{1}{x\pi L}\left(\frac 12+x^2\right)
\ee
with $x\equiv r_0/L$. Notice that $T_c=\sqrt{2}/\pi L$ is the critical temperature, below which no black hole solution exists.

We now have
\be
\frac{r}{L}=\sqrt{\left(\frac{1}2+x^2\right)\cosh \left(\frac{2\beta}{L}\right)-\frac{1}{2}}
\ee
Curiously, in this case the geometry ends at the energy scale
\be
\Lambda^2=\frac{2}{L^2}(1+2x^2)=\frac 12 (2\pi T)^2(1+\sqrt{1-T_c/T}),
\ee
which is lower than the scale set by $T$. Following the reasoning at the end of the previous subsection, we expect correlations to die off not at scale $1/T$, but at a somewhat larger scale. This is confirmed by the fact that in this geometry quasinormal modes have frequencies whose imaginary part scales as $r_0<T$.

\subsection{Thermodynamics of entanglement?}
\label{thermoent}

In the $AdS$/CFT correspondence, a given geometry corresponds to some state in the dual field theory. In this state there is a high degree of entanglement, as evidenced by the extremely large value of two point functions of stress tensor correlators. More generically, this amount of entanglement depends on the scale at which one is probing the system. For instance, a gapped state presents no correlations beyond a distance of order of the inverse mass. Similarly, a deconfined finite temperature system will also present no correlations for distances larger than $1/T$. So clearly the amount of entanglement depends on the scale that we're considering. We need then to describe the state at different scales. A proposal for how to do this on the lattice has been made in references \cite{EntRen,Vidal:2008zz}, and has been named the Multi-Scale Entanglement Renormalzation Ansatz (MERA). In this setup one must not only coarse grain but disentangle nearby degrees of freedom as one performs the renormalization group flow. As we do this, the density matrix describing the system stops being pure and gradually becomes more and more mixed. If the number of degrees of freedom being coarse grained are sufficiently entangled with the remainder of the system, this acts as if we were coupling the system to a finite temperature bath. It is tempting to identify such a temperature with $1/\beta$.

With this interpretation, it is interesting to consider how the phase space changes as a function $\beta$. This piece of information is captured by:
\be
\mathcal E=\frac{1}{d-2}\frac{d \log \Omega}{d\beta}.
\ee
This is the same $\mathcal E$ as in \reef{Ei}, where we have dropped the index $i$ by assuming isotropy. The ``energy'' $\mathcal E$ seems telling us something about how many degrees of freedom are active at a given scale, since it characterizes the relationship between $\beta$ and scale, the latter being given essentially by $\Omega$. For instance, in pure $AdS$ we find
\be
\Omega\simeq e^{(d-2)\beta}\qquad \Rightarrow \qquad   \mathcal E=1/L
\ee
This means that in this case, performing a scaling transformation on $\Omega$ corresponds simply to adding a constant to $\beta$, which  then counts  the number of e-folds , or RG transformations. 


The situation becomes more complicated, but also much more interesting in the AdS-Schwarzschild geometry, where we get
\be
\mathcal E=\frac{1}{L}\tanh(2\beta/L)
\ee
Keeping with our interpretation, we see that at the black hole horizon something peculiar happens, since varying the scale by an infinitesimal leads to a finite variation in $\beta$. This means there must be a very large concentration of degrees of freedom at the horizon. This interpretation is therefore in agreement with our finding of a divergence in $\mathcal N(r)$
at the black hole horizon. 

The quantity $\mathcal E$ can be used to relate the two definitions $\Omega$ and $\oef$. These two are quite different. For instance, in the planar black hole, from $\oef$ we read off that the scale corresponding to the horizon is proportional to $g(r_0)$, and therefore it vanishes; whereas from $\Omega$ we get a finite scale corresponding to the temperature. Using $\mathcal E$ however we can relate the two notions of cut-off:
\bea
\frac{\oef}{\Omega}&=& \left(\frac{\Lambda_{\mbox{\tiny{eff}}}}{\Lambda}\right)^{d-2}, \qquad \Lambda_{\mbox{\tiny{eff}}}=\Lambda \times \sqrt g= \Lambda (L \mathcal E).
\eea
Using this we can write
\be
\mathcal N=\frac{N_{\mbox{\tiny{dof}}}}{(L\mathcal E)^{d-2}},
\ee
which establishes the relation between the $\mathcal N$ function and the proposal of \cite{Padmanabhan:2010xh}.

\section{Discussion}
\label{sec6}
In this paper we have constructed a new object, the $\mathcal N$-function which seems to generalize the concept of a holographic $c$-function to geometries not usually associated with renormalization group flows, namely black holes. We have found that the $\mathcal N$-function can be written as
\be
\mathcal N=\frac{S}{4\pi\oef}
\ee
the ratio of an entropy to an effective phase space volume, the latter being indirectly related to the area of a surface in Planck units, and the former being given by Wald's entropy formula. On domain-wall solutions interpolating between two different $AdS$ spaces with different radius but same dimensionality, holographically interpreted as renormalization group flows, the $\mathcal N$- function provides a holographic $c$-function for Lovelock theories of gravity. On black hole geometries, $\mathcal N$ interpolates between the Euler anomaly at the $AdS$ boundary and a quantity directly proportional to the black hole entropy at the horizon. These results suggest that the $\mathcal N$-function is related to the number of effective degrees of freedom holographically describing a given region of spacetime.

In the context of Lovelock theories of gravity we have found that the equation of motion for the background can be recast as a flow equation for the function $\mathcal N$, \reef{Ngrav} which determines its variation depending on the local gravitational field. The existence of such a simple equation is by all means non-trivial, and this has been our guide in defining $\mathcal N$. Although we have not shown it here, we have checked that in quasi-topological gravity, both cubic \cite{Myers:2010ru,Myers:2010jv,Oliva:2010zd,Oliva:2010eb} and quartic\cite{migquart} , this equation is also satisfied, so this does not seem to be an accident relevant only to Lovelock theories of gravity.

Another interesting class of theories to consider, are new massive gravity and its cousins \cite{Bergshoeff:2009fj,Bergshoeff:2009tb,Sinha:2010ai,Paulos:2010ke}, which exist only in three dimensions. In the following we shall consider as an example the new massive gravity case, although our results are generic. For new massive gravity we have the action 
\be
S=\frac {1}{\lp}\int d^3x\sqrt{-g}\left(R+2+4\lambda \left(R_{ab}^2-\frac 43 R^2\right)\right)
\ee
It is known that such theories support a holographic $c$-function. Also, exact black hole solutions can be found with metric
\be
ds^2=-\frac{r^2 g(r)}{L^2 \ff}\, dt^2+ \frac{L^2 dr^2}{r^2 g(r)}+\frac{r^2}{L^2} dx^2,
\ee
and $g(r)=\ff(1-r_0^2/r^2)$, with $\ff$ satisfying a simple polynomial equation, similar to \reef{adsvacua}. 
We define the $\mathcal N$-function as
\be
\mathcal N=\frac{S}{4\pi\oef}=\frac{1+2 \lambda \ff}{\sqrt{g}}=c \times \frac{\lp}{L} \sqrt{\frac{\ff}{g}}.
\ee
Although for these theories, a simple equation for the flow of $\mathcal N$ is hard to find, it is clear that with the black hole solution we have provided the above is monotonously increasing towards the IR, where it diverges, analogous to our results for the planar black hole in Lovelock gravity. Interestingly, the numerator in the above matches the central charge $c$ of the theory up to a factor of $L/\lp$. The flow of $\mathcal N$ in this case is trivial, since only the denominator is changing. In a sense this is telling us why black hole entropy for three-dimensional theories of gravity is directly determined by the dual CFT central charge - it is because the $\mathcal N$-function is trivial. For higher dimensional gravity theories, such as Lovelock theories, the flow of $\mathcal N$ is non-trivial, and in general there is no direct connection between the value of the Euler anomaly and black hole entropy.

In section 5 we have argued that the natural parameter parameterizing the renormalization group flow from the gravitational point of view, is the proper radial distance $\beta$, and we have shown that gravitational quantities such as curvature seem to have a thermodynamic interpretation when $\beta$ is taken to be an inverse temperature. We have suggested that $\beta$ is possibly related to the fact that if there is sufficient entanglement between low and high scales, the latter act as a finite temperature bath from the point of view of the latter. As far as we know, entanglement entropy calculations haven't been performed in momentum space. While such a calculation yields zero for free field theories, it is non-zero when interactions are turned on. The challenge of course is whether one can say anything useful about it for strongly coupled field theories, and for this $AdS$/CFT should provide clues. It has been suggested that real space entanglement entropy can be evaluated in this context by computing the area of minimal surfaces \cite{Ryu:2006bv,Ryu:2006ef} (see \cite{Nishioka:2009un} for a recent review). Since we know that the radial coordinate in $AdS$ is related to scale, the natural guess for what this momentum space entanglement entropy is, is simply what we have defined as $S$, the Wald entropy formula evaluated on a constant radial surface. It would be interesting to develop this further.


The main unresolved issue in this paper is a proper understanding of the divergence of the $\mathcal N$-function. While for planar black holes, this divergence can be more or less hand-waved away, in global $AdS$ this poses a much tougher problem, which any interpretation of $\mathcal N$ must provide. It is important to notice however, that regardless of any interpretation, the fact that $\mathcal N$ satisfies the simple equation \reef{alg} suggests that it is a quantity worthy of study, even for the global $AdS$ geometries. In other words, this divergence is not an artifact of some artificial definition of $\mathcal N$: rather this definition was arrived at by directly considering the equations of motion, which compels us to take it seriously.

With this in mind, we should ask ourselves what is so special about the point where $g=0$? At this point the $\mathcal N$ function matches that of an equivalent planar black hole. This is because the curvature in the $(d-2)$ spatial directions is given by $-g(r)/L^2$, and is therefore vanishing at that point, beyond which it flips sign, becoming positive. In the spherical black hole, the divergence at finite radius becomes closer and closer to the black hole horizon when we increase the temperature, or cranking up $r_0$. The curvature in the $(d-2)$ spatial directions is given by $-g(r)/L^2$, and at the horizon this is $-1/r_0^2$. Therefore by increasing the temperature, the $g=0$ surface and the horizon coalesce, and $\mathcal N(r_0)$ approaches $\mathcal N(g=0)$, and therefore diverges. So in a sense the divergence at the horizon found in the planar black hole case, can be thought of as the infinite temperature divergence of the spherical case, where the horizon is becoming larger and flatter. 

A possible clue is that at the $g=0$ point, the effective phase space volume $\oef$ vanishes, and beyond this point it becomes imaginary. Using equation \reef{kkeqn}, this is precisely the point where the effective momentum scale equals the gap scale, so it makes sense that $\oef$ vanishes there. If we were in empty $AdS$, the geometry would stop precisely at this point. In the black hole case this is not true, and therefore it is likely that the interpretation of $\oef$ as an effective phase space breaks down. Perhaps it doesn't make sense to talk about $\mathcal N$ beyond the $g=0$ point; or more likely its divergence is signalling some fundamental change in the nature of the holographic degrees of freedom responsible for describing the geometry beyond that scale. 

Finally, the work in this paper suggests that the study of renormalization group flows in non-trivial (i.e. not vacuum) states of a conformal field theory is worthy of study, and presents a small step in this direction. That this is an interesting problem is directly suggested by $AdS$/CFT black hole background solutions. There is by now strong evidence that there is an emergent conformal symmetry at black hole horizons (see e.g. \cite{Carlip:2008rk}), which is realized in full for extremal charged black holes possessing a near horizon $AdS_2$ region. This means that there should be interesting, non-vacuum RG flows between CFT fixed points with different dimensionality. It is possible that the divergence of $\mathcal N$ might be related to this dimensionality reduction, where spatial field modes are turned into a large set of new fields. We leave this and many other questions for future consideration.

\acknowledgments

It is a pleasure to acknowledge discussions and comments from Rob Myers, Aninda Sinha, Jonathan Shock, Javier Tarrio, Ayan Mukhopadhyay, Erik Verlinde, Constantin Bachas and Sheer El-Showk. We acknowledge the hospitality of the Kavli Institute for Theoretical Physics China, where part of this work was performed. This research was supported in part by the Project of Knowledge Innovation Program (PKIP) of Chinese Academy of Sciences, Grant No. KJCX2.YW.W10. 
The author acknowledges funding from the LPTHE, Universit\'e Pierre et Marie Curie.

\appendix

\section{Lovelock theories and black hole solutions}
\label{Lovelock}

\subsection{Lovelock gravity}
In this section we provide a brief review of Lovelock theories of gravity and some of their solutions. We will mostly follow a notation consistent with \cite{Camanho:2010ru}. These theories are the most general second order gravity theories which are also free of ghosts when expanding about flat space \cite{Lovelock1971,Zumino1986}. In terms of differential forms we can write the action
\footnote{We employ differential forms, namely the vierbein $e^a$, and spin connection, $\omega_{~b}^a$, 1-forms, out of which we can construct the Riemann curvature, $R_{~b}^a$, and torsion, $T^a$, 2-forms. This is done via the equation
\begin{equation*}
R_{~b}^a = d\,\omega_{~b}^a + \omega_{~c}^a \wedge \omega_{~b}^c =  \frac{1}{2} R_{~b\mu\nu}^{a}\; dx^{\mu} \wedge dx^{\nu} ~, \qquad T^a = d\,e^a + \omega_{~b}^a \wedge e^b ~.
\end{equation*}}
 as
\begin{equation}
I_g = \frac{1}{\lp^{d-2} L^2}\sum_{k=0}^{K} {\frac{c_k}{(d-2k)(d-3)!}} \mathop\int \mathcal{R}^{(k)} ~.
\label{LLaction}
\end{equation}
Here $L$ is some length scale related to the cosmological constant, $c_k$ is an arbitrary set of dimensionless couplings and a $K$ is a positive integer restricted to $K\leq \left[\frac{d-1}{2}\right]$. $\mathcal{R}^{(k)}$ is the exterior product of $k$ curvature 2-forms with the required number of vierbeins to construct a $d$-form, 
\begin{equation}
\mathcal{R}^{(k)} = \epsilon_{f_1 \cdots f_{d}}\; R^{f_1 f_2} \wedge \cdots \wedge R^{f_{2k-1} f_{2k}} \wedge e^{f_{2k+1} \cdots f_d} ~,
\end{equation}
where $e^{{f_1}\cdots{f_k}}$ is a short notation for $e^{f_1} \wedge \ldots \wedge e^{f_k}$. The zeroth and first term in (\ref{LLaction}) correspond, respectively, to the cosmological constant and the Einstein-Hilbert lagrangian, with $c_0=c_1=1$. This choice implies the cosmological constant is given by $\Lambda=-(d-1)(d-2)/L^2$.

In the first order formalism, the action leads to two equations of motion,  for the connection 1-form and for the vierbein. The variation of the action with respect to the connection gives an equation which is proportional to torsion, which we set to zero. By varying with respect to the vierbein we obtain
\be
l_p^{d-2} L^2 (d-3)! E_a=\epsilon_{a f_1\ldots f_{d-1}} \sum_k c_k \, R^{f_1 f_2}\wedge \ldots \wedge R^{f_{2k-1} f_{2k}}\wedge e^{f_{2k+1}\ldots f_{d-1}}=0. \label{ealove}
\ee

\subsection{Black hole solutions}

We take the following metric ansatz
\be
ds^2=-(\kappa+F(r)) dt^2+\frac{L^2 dr^2}{\kappa+G(r)}+\frac{r^2}{L^2} (\ud \Sigma^{d-2}_\kappa)^2 \label{bgappendix}
\ee
with $\ud \Sigma$ is the volume element for the space with unit radius and constant curvature $\kappa=-1,0,1$. 
For the vierbein we choose
\be
e^{\hat t}=\sqrt{F(r)} \ud t, \qquad e^{\hat r}=\frac{1}{\sqrt{G(r)}}\,\ud r, \qquad e^{\hat i}=\frac{r}{L} \epsilon^i
\ee
where hatted indices indicate flat coordinates. The $\epsilon^i$ describe a vierbein for the constant curvature space $\Sigma$. In these coordinates the spin connection and curvature two forms are easily computed. We obtain
\bea
R^{\hat t \hat r}&=& \left(\frac{(\kappa+G) F'^2}{4 (\kappa+F)^2}-\frac{F' G'}{4(\kappa+F)}-\frac 12 \frac{(\kappa+G) F''}{\kappa+F}\right) e^{\hat t}\wedge e^{\hat r} \nonumber \\
R^{\hat r \hat i}&=& -\frac{L G'}{2r}\, e^{\hat r}\wedge e^{\hat i}, \nonumber \\
R^{\hat t \hat i}&=& -\frac{L(\kappa+G) F'}{2r (\kappa+F)}\, e^{\hat t}\wedge e^{\hat i},\nonumber \\
R^{\hat i \hat j}&=& - \frac{L^2 G}{r^2}\, e^{\hat i}\wedge e^{\hat j} \label{curvs}
\eea
Going back to the equation of motion \reef{ealove} and looking at the $t$ component we obtain
\bea
L^2 \lp^{d-2}(d-3)! E_t&=& \sum_k c_k (d-2)!\left[ 2k\, R^{\hat r \hat i}_{\ \ \hat r \hat i} \left(R^{\hat i \hat j}_{\ \ \hat i \hat j}\right)^{k-1}+(d-2k-1) \left(R^{\hat i \hat j}_{\ \ \hat i \hat j}\right)^k\right]\nonumber \\
&=&\sum_k (-1)^k\,c_k \frac{(d-2)!}{2}\, r^{-(d-2)}\, \frac{\ud}{\ud r}\left[r^{d-1} \left(\frac{G(r) L^2}{r^2}\right)^k\right]
\eea
Defining $G(r)=\frac{r^2}{L^2} g(r)$ and
\be
\Upsilon[g]\equiv \sum_k (-1)^k c_k  g^k
\ee
we obtain
\bea
E_t&=&\frac{(d-2)}{2 L^2 \lp^{d-2}}\left[(d-1)\Upsilon[g]+\Upsilon'[g]\, r g'\right] \nonumber \\
&=& \frac{(d-2)}{2 L^2 \lp^{d-2}} r^{-(d-2)}\, \frac{\ud}{\ud r} \left( r^{d-1} \Upsilon[g]\right).
\eea
The $r$ equation is similarly deduced. We get
\bea
L^2 \lp^{d-2}(d-3)! E_r&=& \sum_k c_k (d-2)!\left[ 2k\, R^{\hat t \hat i}_{\ \ \hat t \hat i} \left(R^{\hat i \hat j}_{\ \ \hat i \hat j}\right)^{k-1}+(d-2k-1) \left(R^{\hat i \hat j}_{\ \ \hat i \hat j}\right)^k\right].
\eea
Using expressions \reef{curvs} and replacing $F(r)=\frac{r^2}{L^2} f(r)$ one easily finds
\be
E_r=\frac{(d-2)}{2 L^2 \lp^{d-2}}\,\bigg[(d-1) \Upsilon[g]+\Upsilon'[g]\, 
\left(\frac{\kappa\, L^2+r^2\,g }{\kappa\, L^2+r^2\,f}\, r f'
+\kappa \frac{f-g}{\kappa\, L^2+r^2\,f}\right)
\bigg].
\ee

Taking the difference of the two equations we find
\be
E_t-E_r=\frac{(d-2)}{2\lp^{d-2}}\frac{ \Upsilon'[g]}{\kappa L^2+r^2 f}\left[\frac {\kappa L^2} r\left(r^2 f-r^2 g\right)'+r^3 (g f'-f g')
\right].
\ee
This implies that in the absence of matter or for matter saturating the null energy condition, we must have $g(r)=f(r)$. 

\subsection{Evaluation of the Wald formula}
\label{waldform}
The Wald formula is given by
\be
S=-2\pi \int \sqrt{h} \frac{\partial \mathcal L}{\partial R_{abcd}}\epsilon_{ab}\epsilon_{cd} \label{s1}
\ee
On the background \reef{bgappendix} we choose constant $r,t$ surfaces, so that the surface binormals $\epsilon_{ab}$ only have non-zero components $\epsilon_{rt}=\sqrt{-g_{rr}g_{tt}}$. We then get
\be
S=\frac{8\pi}{\lp^{d-2}}\, V_{d-2} \left(\frac{r}{L}\right)^{d-2} \frac{\partial \mathcal L}{\partial R_{rt}^{\ \ rt}}
\ee
Going back to the action \reef{LLaction}, we write it in tensor notation:
\be
I_g=\frac{1}{L^2 \lp^{d-2}}\sum_{k=0}^K\int d^d x \sqrt{-g}\left(n_k\, c_k\, \delta^{a_1\ldots a_{2k}}_{b_1\ldots b_{2k}}\, R_{a_1 a_2}^{b_1 b_2}\ldots R_{a_{2k-1} a_{2k}}^{b_{2k-1}\, b_{2k}}\right),
\ee
with $n_k=\frac{(d-2k-1)!}{2^k(d-3)!}$. With this action we find
\be
\frac{\partial \mathcal L}{\partial R_{rt}^{\ \ rt}}=\frac{1}{L^2\lp^{d-2}}\sum_{k=1}^K k\,n_k\, c_k \, 
\delta^{i_1\ldots i_{2k-2}}_{j_1\ldots j_{2k-2}}\, R_{i_1 i_2}^{j_1 j_2}\ldots R_{i_{2k-3} i_{2k-2}}^{j_{2k-3}\, j_{2k-2}}
\ee
Since in the spatial directions $i_k$ we have
\be
R^{i_1 i_2}_{\ \ j_1 j_2}=-\frac{g(r)}{L^2}\left(\delta^{i_1}_{j_1}\delta^{i_2}_{j_2}-\delta^{i_2}_{j_1}\delta^{i_1}_{j_2}\right)
\ee
and using the identity ($i_k$ indices run over $(d-2)$ values):
\be
\delta^{i_1\ldots i_k i_{k+1}\ldots i_n}_{i_1\ldots i_k j_{k+1}\ldots j_n}
=\frac{((d-2)-n+k)!}{((d-2)-n)!}\, \delta^{i_{k+1}\ldots i_n}_{j_{k+1}\ldots j_n}
\ee
we get
\be
\frac{\partial \mathcal L}{\partial R_{rt}^{\ \ rt}}=\frac{1}{2 L^2\lp^{d-2}}\sum_{k=1}^K \frac{(d-2)k}{d-2k}c_k (-g)^{k-1} \label{llwald}
\ee

\section{A Wald formula for the Euler anomaly for generic theories of gravity}
\label{Euler}
Consider a general gravity theory supporting an $AdS$ vacua in $d$ dimensions, with $d$ odd. The holographically dual field theory is conformal on a flat metric, but anomalies break conformal invariance on a curved background. Generically such an anomaly is of the form
\be
\langle T^{a}_{a}\rangle= (-1)^{D/2}\, A E_{d-1}+\sum b_k C^k+\mbox{total derivatives}.
\ee
Here $E_{d-1}$ is the $d-1$-dimensional Euler density and $C^k$ are some contractions of $k$ Weyl tensors. The computation of these anomalies on the gravitational theory was first accomplished by \cite{Henningson:1998gx,Henningson:1998ey}.
In \cite{Imbimbo:1999bj} it was shown that the Euler anomaly as given by the $A$ coefficient, can be simply computed by evaluating the gravitational lagrangian on-shell on an $AdS$ background and considering the coefficient of the leading divergent piece. Here we shall show how this is related to the Wald formula.

We consider a general higher derivative gravity theory with action 
\bea
S&=&\int d^{d}x\sqrt{-g}(\mathcal L_g+\mathcal L_m),\nonumber \\
\mathcal L_g&=& \sum_k L^{(k)}.
\eea
Here $L_m$ is the matter lagrangian and we decompose the gravitational lagrangian $\mathcal L_g$ according to the number of curvatures each term possesses, so that $L_1$ has one curvature, $L_2$ has two, etc.
The above action leads to the equation of motion
\be
-2 \nabla_a \nabla_b X^{acbd}+X^{aefc}R_{aef}^{\ \ \ d}+\frac 12 g^{cd}\mathcal L_g+ \frac{\partial \mathcal L}{\partial g_{cd}}=T^{cd}
\ee
where $T^{cd}$ is the matter sector stress tensor and
\be
X^{abcd}=\frac{\delta S}{\delta R_{abcd}}.
\ee 
Now assume we have an $AdS$ background with some radius,
\be
ds^2=\frac{L^2\ud r^2}{\ff r^2}+\frac{r^2}{L^2}(-\ud t^2+\ud \mbf x^2)
\ee
In these circumstances we must have
$T^{cd}=-\frac 12 g^{cd} \mathcal L_m$, and all covariant derivatives vanish. Taking the trace of the equation of motion we find
\bea
\sum_k \left(k L_k+\frac d2 L_k-2k L_k)\right)+\frac d2 \mathcal L_m &=& 0  
\eea
This implies
\be
X^{abcd}R_{abcd}=\sum k L_k=\frac d2 \left(\mathcal L_g+\mathcal L_m)\right.
\ee
This tells us that the on-shell lagrangian on $AdS$ is related to $X^{abcd}$. We also have
\bea
X^{abcd}=(g^{ac}g^{bd}-g^{bc}g^{ad}) X^{rt}_{\ \ rt},
\eea
and therefore we conclude
\bea
\mathcal L_g+\mathcal L_m &=& \frac 4d R \frac{\delta S}{\delta R^{rt}_{\ \ rt}}
\eea
This is very similar to Wald's entropy formula. In fact, it is now straightforward to verify that the $A$ anomaly coefficient (with a suitable normalization) is given by
\be
A=\frac 12 \frac{\partial \mathcal L_g}{\partial R_{abcd}}\, \epsilon_{ab}\epsilon_{cd}
\ee
where $\epsilon_{rt}=\sqrt{-g_{rr}g_{tt}}$ with all other components zero is a spacelike surface binormal.

In particular, for Lovelock theories of gravity, we can use equation \reef{llwald} to get
\be
A=\left(\frac{L}{\lp}\right)^{d-2}\frac{\sum_k \frac{(d-2) k}{d-2k}\,c_k (-\ff)^{k-1}}{\ff^{(d-2)/2}}.
\ee
This result is obtained by picking the Poincar\'e patch of $AdS$ as the background: in \reef{bgappendix} we take $\kappa=0$, $F=\frac{r^2}{L^2}$, $G=\frac{r^2}{L^2} \ff$, corresponding to an $AdS$ space of effective radius $L/\sqrt{\ff}$.

\bibliography{Biblio}{}

\providecommand{\href}[2]{#2}\begingroup\raggedright\begin{thebibliography}{10}

\bibitem{Zamolodchikov:1986gt}
A.~B. Zamolodchikov, {\it {Irreversibility of the Flux of the Renormalization
  Group in a 2D Field Theory}},  {\em JETP Lett.} {\bf 43} (1986) 730--732.

\bibitem{Cardy:1988cwa}
J.~L. Cardy, {\it {Is There a c Theorem in Four-Dimensions?}},  {\em Phys.
  Lett.} {\bf B215} (1988) 749--752.

\bibitem{Maldacena:1997re}
J.~M. Maldacena, {\it {The large N limit of superconformal field theories and
  supergravity}},  {\em Adv. Theor. Math. Phys.} {\bf 2} (1998) 231--252,
  [\href{http://arxiv.org/abs/hep-th/9711200}{{\tt hep-th/9711200}}].

\bibitem{Aharony:1999ti}
O.~Aharony, S.~S. Gubser, J.~M. Maldacena, H.~Ooguri, and Y.~Oz, {\it {Large N
  field theories, string theory and gravity}},  {\em Phys. Rept.} {\bf 323}
  (2000) 183--386, [\href{http://arxiv.org/abs/hep-th/9905111}{{\tt
  hep-th/9905111}}].

\bibitem{Girardello:1998pd}
L.~Girardello, M.~Petrini, M.~Porrati, and A.~Zaffaroni, {\it {Novel local CFT
  and exact results on perturbations of N=4 superYang Mills from AdS
  dynamics}},  {\em JHEP} {\bf 9812} (1998) 022,
  [\href{http://arxiv.org/abs/hep-th/9810126}{{\tt hep-th/9810126}}].

\bibitem{Freedman:1999gp}
D.~Z. Freedman, S.~S. Gubser, K.~Pilch, and N.~P. Warner, {\it {Renormalization
  group flows from holography supersymmetry and a c-theorem}},  {\em Adv.
  Theor. Math. Phys.} {\bf 3} (1999) 363--417,
  [\href{http://arxiv.org/abs/hep-th/9904017}{{\tt hep-th/9904017}}].

\bibitem{Nojiri:1999mh}
S.~Nojiri and S.~D. Odintsov, {\it {On the conformal anomaly from higher
  derivative gravity in AdS/CFT correspondence}},  {\em Int. J. Mod. Phys.}
  {\bf A15} (2000) 413--428, [\href{http://arxiv.org/abs/hep-th/9903033}{{\tt
  hep-th/9903033}}].

\bibitem{Sinha:2010ai}
A.~Sinha, {\it {On the new massive gravity and AdS/CFT}},
  \href{http://arxiv.org/abs/1003.0683}{{\tt arXiv:1003.0683}}.

\bibitem{Sinha:2010pm}
A.~Sinha, {\it {On higher derivative gravity, c-theorems and cosmology}},
  \href{http://arxiv.org/abs/1008.4315}{{\tt arXiv:1008.4315}}.

\bibitem{Paulos:2010ke}
M.~F. Paulos, {\it {New massive gravity extended with an arbitrary number of
  curvature corrections}},  {\em Phys. Rev.} {\bf D82} (2010) 084042,
  [\href{http://arxiv.org/abs/1005.1646}{{\tt arXiv:1005.1646}}].

\bibitem{Myers:2010xs}
R.~C. Myers and A.~Sinha, {\it {Seeing a c-theorem with holography}},  {\em
  Phys. Rev.} {\bf D82} (2010) 046006,
  [\href{http://arxiv.org/abs/1006.1263}{{\tt arXiv:1006.1263}}].

\bibitem{Susskind:1998dq}
L.~Susskind and E.~Witten, {\it {The holographic bound in anti-de Sitter
  space}},  \href{http://arxiv.org/abs/hep-th/9805114}{{\tt hep-th/9805114}}.

\bibitem{Padmanabhan:2010xh}
T.~Padmanabhan, {\it {Surface Density of Spacetime Degrees of Freedom from
  Equipartition Law in theories of Gravity}},  {\em Phys. Rev.} {\bf D81}
  (2010) 124040, [\href{http://arxiv.org/abs/1003.5665}{{\tt
  arXiv:1003.5665}}].

\bibitem{Iyer:1994ys}
V.~Iyer and R.~M. Wald, {\it {Some properties of Noether charge and a proposal
  for dynamical black hole entropy}},  {\em Phys. Rev.} {\bf D50} (1994)
  846--864, [\href{http://arxiv.org/abs/gr-qc/9403028}{{\tt gr-qc/9403028}}].

\bibitem{Wald:1993nt}
R.~M. Wald, {\it {Black hole entropy is the Noether charge}},  {\em Phys. Rev.}
  {\bf D48} (1993) 3427--3431, [\href{http://arxiv.org/abs/gr-qc/9307038}{{\tt
  gr-qc/9307038}}].

\bibitem{Faulkner:2009wj}
T.~Faulkner, H.~Liu, J.~McGreevy, and D.~Vegh, {\it {Emergent quantum
  criticality, Fermi surfaces, and AdS2}},
  \href{http://arxiv.org/abs/0907.2694}{{\tt arXiv:0907.2694}}.

\bibitem{Solodukhin:1998tc}
S.~N. Solodukhin, {\it {Conformal description of horizon's states}},  {\em
  Phys. Lett.} {\bf B454} (1999) 213--222,
  [\href{http://arxiv.org/abs/hep-th/9812056}{{\tt hep-th/9812056}}].

\bibitem{Carlip:2008rk}
S.~Carlip, {\it {Black Hole Entropy and the Problem of Universality}},
  \href{http://arxiv.org/abs/0807.4192}{{\tt arXiv:0807.4192}}.

\bibitem{Lovelock1971}
D.~Lovelock, {\it {The Einstein tensor and its generalizations}},  {\em J.
  Math. Phys.} {\bf 12} (1971) 498--501.

\bibitem{Zumino1986}
B.~Zumino, {\it {Gravity Theories in More Than Four-Dimensions}},  {\em Phys.
  Rept.} {\bf 137} (1986) 109.

\bibitem{Hofman:2009ug}
D.~M. Hofman, {\it {Higher Derivative Gravity, Causality and Positivity of
  Energy in a UV complete QFT}},  {\em Nucl. Phys.} {\bf B823} (2009) 174--194,
  [\href{http://arxiv.org/abs/0907.1625}{{\tt arXiv:0907.1625}}].

\bibitem{Buchel:2009sk}
A.~Buchel {\em et~al.}, {\it {Holographic GB gravity in arbitrary dimensions}},
   {\em JHEP} {\bf 03} (2010) 111, [\href{http://arxiv.org/abs/0911.4257}{{\tt
  arXiv:0911.4257}}].

\bibitem{Camanho:2010ru}
X.~O. Camanho, J.~D. Edelstein, and M.~F. Paulos, {\it {Lovelock theories,
  holography and the fate of the viscosity bound}},
  \href{http://arxiv.org/abs/1010.1682}{{\tt arXiv:1010.1682}}.

\bibitem{Heemskerk:2010hk}
I.~Heemskerk and J.~Polchinski, {\it {Holographic and Wilsonian Renormalization
  Groups}},  \href{http://arxiv.org/abs/1010.1264}{{\tt arXiv:1010.1264}}.

\bibitem{Verlinde:2010hp}
E.~P. Verlinde, {\it {On the Origin of Gravity and the Laws of Newton}},
  \href{http://arxiv.org/abs/1001.0785}{{\tt arXiv:1001.0785}}.

\bibitem{Myers:2010tj}
R.~C. Myers and A.~Sinha, {\it {Holographic c-theorems in arbitrary
  dimensions}},  \href{http://arxiv.org/abs/1011.5819}{{\tt arXiv:1011.5819}}.

\bibitem{Liu:2010xc}
J.~T. Liu, W.~Sabra, and Z.~Zhao, {\it {Holographic c-theorems and higher
  derivative gravity}},  \href{http://arxiv.org/abs/1012.3382}{{\tt
  arXiv:1012.3382}}.

\bibitem{Cai:1998vy}
R.-G. Cai and K.-S. Soh, {\it {Topological black holes in the dimensionally
  continued gravity}},  {\em Phys. Rev.} {\bf D59} (1999) 044013,
  [\href{http://arxiv.org/abs/gr-qc/9808067}{{\tt gr-qc/9808067}}].

\bibitem{Cai:2001dz}
R.-G. Cai, {\it {Gauss-Bonnet black holes in AdS spaces}},  {\em Phys. Rev.}
  {\bf D65} (2002) 084014, [\href{http://arxiv.org/abs/hep-th/0109133}{{\tt
  hep-th/0109133}}].

\bibitem{Cai:2003kt}
R.-G. Cai, {\it {A note on thermodynamics of black holes in Lovelock gravity}},
   {\em Phys. Lett.} {\bf B582} (2004) 237--242,
  [\href{http://arxiv.org/abs/hep-th/0311240}{{\tt hep-th/0311240}}].

\bibitem{Jacobson:1993vj}
T.~Jacobson, G.~Kang, and R.~C. Myers, {\it {On black hole entropy}},  {\em
  Phys. Rev.} {\bf D49} (1994) 6587--6598,
  [\href{http://arxiv.org/abs/gr-qc/9312023}{{\tt gr-qc/9312023}}].

\bibitem{Boulware:1985wk}
D.~G. Boulware and S.~Deser, {\it {String Generated Gravity Models}},  {\em
  Phys. Rev. Lett.} {\bf 55} (1985) 2656.

\bibitem{Camanho:2009hu}
X.~O. Camanho and J.~D. Edelstein, {\it {Causality in AdS/CFT and Lovelock
  theory}},  \href{http://arxiv.org/abs/0912.1944}{{\tt arXiv:0912.1944}}.

\bibitem{deBoer:2009pn}
J.~de~Boer, M.~Kulaxizi, and A.~Parnachev, {\it {AdS7/CFT6, Gauss-Bonnet
  Gravity, and Viscosity Bound}},  \href{http://arxiv.org/abs/0910.5347}{{\tt
  arXiv:0910.5347}}.

\bibitem{LargeScale}
S.~Hawking and G.~Ellis, {\em {The Large scale structure of space-time}}.
\newblock {Cambridge University Press}, Cambridge, UK, first paperback edition
  with corrections~ed., 1973.
\newblock Cambridge Monographs On Mathematical Physics.

\bibitem{'tHooft:1993gx}
G.~'t~Hooft, {\it {Dimensional reduction in quantum gravity}},
  \href{http://arxiv.org/abs/gr-qc/9310026}{{\tt gr-qc/9310026}}.

\bibitem{BoschiFilho:1994an}
H.~Boschi-Filho, C.~Farina, and A.~de~Souza~Dutra, {\it {The Partition function
  for an anyon - like oscillator}},  {\em J. Phys.} {\bf A28} (1995) L7--L12,
  [\href{http://arxiv.org/abs/hep-th/9410098}{{\tt hep-th/9410098}}].

\bibitem{EntRen}
G.~{Vidal}, {\it {Entanglement Renormalization}},  {\em Physical Review
  Letters} {\bf 99} (Nov., 2007) 220405--+,
  [\href{http://arxiv.org/abs/arXiv:cond-mat/0512165}{{\tt
  arXiv:cond-mat/0512165}}].

\bibitem{Vidal:2008zz}
G.~Vidal, {\it {Class of Quantum Many-Body States That Can Be Efficiently
  Simulated}},  {\em Phys. Rev. Lett.} {\bf 101} (2008) 110501.

\bibitem{Myers:2010ru}
R.~C. Myers and B.~Robinson, {\it {Black Holes in Quasi-topological Gravity}},
  {\em JHEP} {\bf 08} (2010) 067, [\href{http://arxiv.org/abs/1003.5357}{{\tt
  arXiv:1003.5357}}].

\bibitem{Myers:2010jv}
R.~C. Myers, M.~F. Paulos, and A.~Sinha, {\it {Holographic studies of
  quasi-topological gravity}},  \href{http://arxiv.org/abs/1004.2055}{{\tt
  arXiv:1004.2055}}.

\bibitem{Oliva:2010zd}
J.~Oliva and S.~Ray, {\it {Classification of Six Derivative Lagrangians of
  Gravity and Static Spherically Symmetric Solutions}},  {\em Phys. Rev.} {\bf
  D82} (2010) 124030, [\href{http://arxiv.org/abs/1004.0737}{{\tt
  arXiv:1004.0737}}].

\bibitem{Oliva:2010eb}
J.~Oliva and S.~Ray, {\it {A new cubic theory of gravity in five dimensions:
  Black hole, Birkhoff's theorem and C-function}},  {\em Class. Quant. Grav.}
  {\bf 27} (2010) 225002, [\href{http://arxiv.org/abs/1003.4773}{{\tt
  arXiv:1003.4773}}].

\bibitem{migquart}
M.~Paulos, {\it {Work in progress}}, .

\bibitem{Bergshoeff:2009fj}
E.~Bergshoeff, O.~Hohm, and P.~Townsend, {\it {On massive gravitons in 2+1
  dimensions}},  \href{http://arxiv.org/abs/0912.2944}{{\tt arXiv:0912.2944}}.

\bibitem{Bergshoeff:2009tb}
E.~A. Bergshoeff, O.~Hohm, and P.~K. Townsend, {\it {On Higher Derivatives in
  3D Gravity and Higher Spin Gauge Theories}},  {\em Annals Phys.} {\bf 325}
  (2010) 1118--1134, [\href{http://arxiv.org/abs/0911.3061}{{\tt
  arXiv:0911.3061}}].

\bibitem{Ryu:2006bv}
S.~Ryu and T.~Takayanagi, {\it {Holographic derivation of entanglement entropy
  from AdS/CFT}},  {\em Phys. Rev. Lett.} {\bf 96} (2006) 181602,
  [\href{http://arxiv.org/abs/hep-th/0603001}{{\tt hep-th/0603001}}].

\bibitem{Ryu:2006ef}
S.~Ryu and T.~Takayanagi, {\it {Aspects of holographic entanglement entropy}},
  {\em JHEP} {\bf 08} (2006) 045,
  [\href{http://arxiv.org/abs/hep-th/0605073}{{\tt hep-th/0605073}}].

\bibitem{Nishioka:2009un}
T.~Nishioka, S.~Ryu, and T.~Takayanagi, {\it {Holographic Entanglement Entropy:
  An Overview}},  {\em J. Phys.} {\bf A42} (2009) 504008,
  [\href{http://arxiv.org/abs/0905.0932}{{\tt arXiv:0905.0932}}].

\bibitem{Henningson:1998gx}
M.~Henningson and K.~Skenderis, {\it {The holographic Weyl anomaly}},  {\em
  JHEP} {\bf 07} (1998) 023, [\href{http://arxiv.org/abs/hep-th/9806087}{{\tt
  hep-th/9806087}}].

\bibitem{Henningson:1998ey}
M.~Henningson and K.~Skenderis, {\it {Holography and the Weyl anomaly}},  {\em
  Fortsch. Phys.} {\bf 48} (2000) 125--128,
  [\href{http://arxiv.org/abs/hep-th/9812032}{{\tt hep-th/9812032}}].

\bibitem{Imbimbo:1999bj}
C.~Imbimbo, A.~Schwimmer, S.~Theisen, and S.~Yankielowicz, {\it
  {Diffeomorphisms and holographic anomalies}},  {\em Class. Quant. Grav.} {\bf
  17} (2000) 1129--1138, [\href{http://arxiv.org/abs/hep-th/9910267}{{\tt
  hep-th/9910267}}].

\end{thebibliography}\endgroup
\bibliographystyle{JHEP}

\end{document}